\documentclass[letterpaper,12pt]{article}
\usepackage{amssymb}
\usepackage{graphicx}
\usepackage{anysize}

\def\beq{\begin{equation}}

\def\eeq{\end{equation}}

\def\lphi{\Lambda_\phi}
\def\mh{m_h}
\def\mphi{m_\phi}

\def\non{\nonumber}

\def\gev{\, {\rm GeV}}

\def\bea{\begin{eqnarray}}

\def\eea{\end{eqnarray}}

\def\beq{\begin{equation}}

\def\eeq{\end{equation}}

\newcommand{\gsim}{\lower.7ex\hbox{$\;\stackrel{\textstyle>}{\sim}\;$}}
\newcommand{\lsim}{\lower.7ex\hbox{$\;\stackrel{\textstyle<}{\sim}\;$}}

\begin{document}


\begin{center}
  \begin{Large}
    \begin{bf}

Precision Electroweak Constraints and the Mixed Radion-Higgs Sector\footnote{\uppercase{T}alk presented at {\it \uppercase{SUSY} 2003:
 \uppercase{S}upersymmetry in the \uppercase{D}esert}\/, 
 held at the \uppercase{U}niversity of \uppercase{A}rizona,
 \uppercase{T}ucson, \uppercase{AZ}, \uppercase{J}une 5-10, 2003.
 \uppercase{T}o appear in the \uppercase{P}roceedings.}
    \end{bf}
  \end{Large}
\end{center}
\vspace{0.1cm}
\begin{center}
{\bf Manuel Toharia\\}
  \vspace{0.1cm}
{ \it Physics Department, University of California, Davis, CA 95616  \\
E-mail: mtoharia@physics.ucdavis.edu}
\end{center}
\vspace{0.4cm}





\abstract{
Adding radion perturbations (up to second order) to the static (RS) metric 
allows us to calculate the general first and second order interactions 
of the radion field with the electroweak vector bosons. We use these interactions to compute 
precision electroweak observables in the case of Higgs-radion mixing and compare with experiment.
}

\vspace{0.4cm}

\section{Introduction}

An interesting Brane World Scenario proposed by Randall and Sundrum (RS)\cite{Randall:1999ee} 
involves one extra dimension with warped geometry. This warping can account for the hierarchy between
the weak scale and the Planck scale.
In this framework the corrections to precision electroweak observables can
be described by the Peskin-Takeuchi parameters $S$ and $T$, and can
come both from the Higgs-radion sector\cite{Csaki:2000zn,Kribs:2001ic,Das:2001pn} and from the Kaluza Klein
excitation sector\cite{Davoudiasl:2000wi}. Here we would like to refine 
and expand upon precision electroweak constraints on the
Higgs-radion sector\cite{pewradion}, focusing on those regions of parameter space for
which the KK excitations are too massive to have significant $S$ and
$T$ contributions.

Ignoring tensor perturbations, the metric of the five-dimensional space can be written up 
to quadratic terms in the radion $r(x)$ as 
\bea ds^2
&=& \left[ e^{-2\sigma}\eta_{\mu\nu} - \hat{\kappa}\left\{\eta_{\mu\nu}\ c(y) r(x)\right\} - \hat{\kappa}^2
  \eta_{\mu\nu} e^{2\sigma} a(y) r^2(x)   \right]  dx^\mu dx^\nu \non\\
&& \hspace{1.3cm} + \ \ \ \left[1 + \hat{\kappa} 2 e^{2\sigma} e(y) r(x)
  + \hat{\kappa}^2 e^{4\sigma}f(y) r^2(x) \right] dy^2 \eea 
where $\hat\kappa^2=M_{\rm Pl_5}^{-3}$ and $r(x)$ is the non-canonically
normalized radion field.
In the unphysical case of no radion stabilization
(i.e., no radion mass) $\sigma(y)=ky$ (in the bulk), $c(y)=e(y)=1$, $a(y)=1/4$ 
and $f(y)=2$ (no $r \partial_\mu r\partial^\mu r$ trilinear self interactions in this limit). 
We assume that all the Standard Model particles live on the ``SM brane'' at $y=y_0$. 
On the other hand, the graviton wave function is peaked at the ``Planck brane'' at $y=0$.

The interactions of the canonically normalized radion $\phi_0$ with the SM vectors, up to order $\phi^2_0$, are
\beq
{{L}}_{int}( \hat{\kappa}^1)=-\left({1\over \lphi}\right)\ 
\phi_0 \left[ M^2_V  V^\alpha V_\alpha - \epsilon \ {L}_V \right]\label{trilag} 
\eeq
\bea
{{L}}_{int}( \hat{\kappa}^2) = {4\over \lphi^2}\ 
\phi_0^2 \left[{\eta\over2} M^2_V  V^\alpha V_\alpha - \epsilon 
\left({(2 \eta -1) \over 4} L_V + {1\over 4} M^2_V  V^\alpha V_\alpha \right) \right]\label{quartlag}
\eea
where $L_V$ is the usual Lagrangian density of a massive vector field. 
$\lphi$ sets the strength of these interactions, 
and should be of order $\simeq 1$ TeV.   
We have also defined $\eta=a(y_0)/c(y_0)^2$ which should be of order unity.
The $\epsilon$ in the above equations is a remnant of the dimensional
regularization requirement of working in $D=4-\epsilon$ dimensions.

There may exist a Higgs-radion mixing\cite{Giudice:2000av} 
term ${ L}= \xi \partial h_0\partial \phi_0$
derived from the SM-brane operator 
\beq
{ L}=\xi \int d^4x \sqrt{g_{\rm ind}} R(g_{\rm ind})H^\dagger H.
\label{higgs radion mixing}
\eeq

The mass eigenstates $h$ and $\phi$ can be obtained from the
``geometry eigenstates'' $h_0$ and $\phi_0$ by suitable field redefinitions\cite{Csaki:2000zn,Dominici:2002jv}.
The interactions between these new physical fields and the electroweak vector bosons will be altered by this mixing.

\section{Precision electroweak constraints}
We can now compute the $S$ and $T$ parameters, both of which have several types of contributions.
First, we have direct contributions of
each eigenstate of the Higgs-radion system.
There are also contributions from the so-called ``anomalous terms'' coming from the 
$\epsilon$ terms in (\ref{trilag}) and (\ref{quartlag}).
Finally, we also expect non-renormalizable operators to contribute. Naively one could expect these operators 
to be suppressed by powers of $M_{Pl}^{-1}$. However as a consequence of the warped geometry, the actual suppression 
is by powers of $\lphi^{-1}$\footnote{To account for these effects, we basically follow Ref\cite{Csaki:2000zn}}.

\begin{figure}[h]
\centering
\includegraphics*[width=6.5cm]{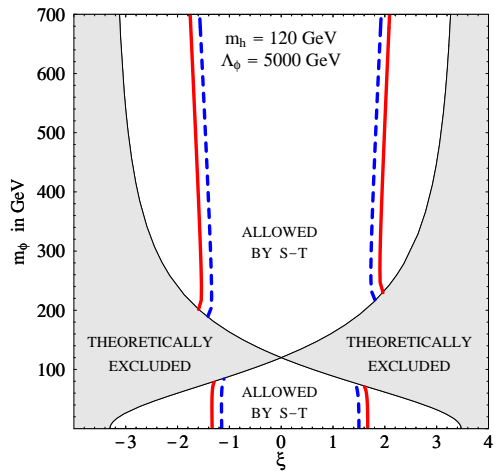}
\caption{Allowed and disallowed regions in $(\xi,\mphi)$
parameter space, assuming $\lphi=5$~TeV, $\mh=120$~GeV 
and $\eta=1/4$. Regions within the theoretically
allowed hourglass shape that are excluded 
at the 68\%  and 90\% confidence level
are those with $|\xi|$ values larger than given by the 
thick dashed (blue) and solid (red) curves, respectively.}
\label{mxi120}
\end{figure}

As shown in Fig.~\ref{mxi120}, when the physical eigenstate masses $\mh$ and
$\mphi$ are both relatively modest in size, precision electroweak constraints disfavor\footnote{In order to determine the allowed regions of $S,T$ parameter space, we have employed a $\chi^2$ ellipse parameterization.} 
the large-$|\xi|$ wings of the theoretically allowed hourglass-shaped
region in the $(\xi,\mphi)$ plane. The dependance on the parameter $\eta$ is minimal.

\begin{figure}[h]
\centering
\includegraphics*[width=6.5cm]{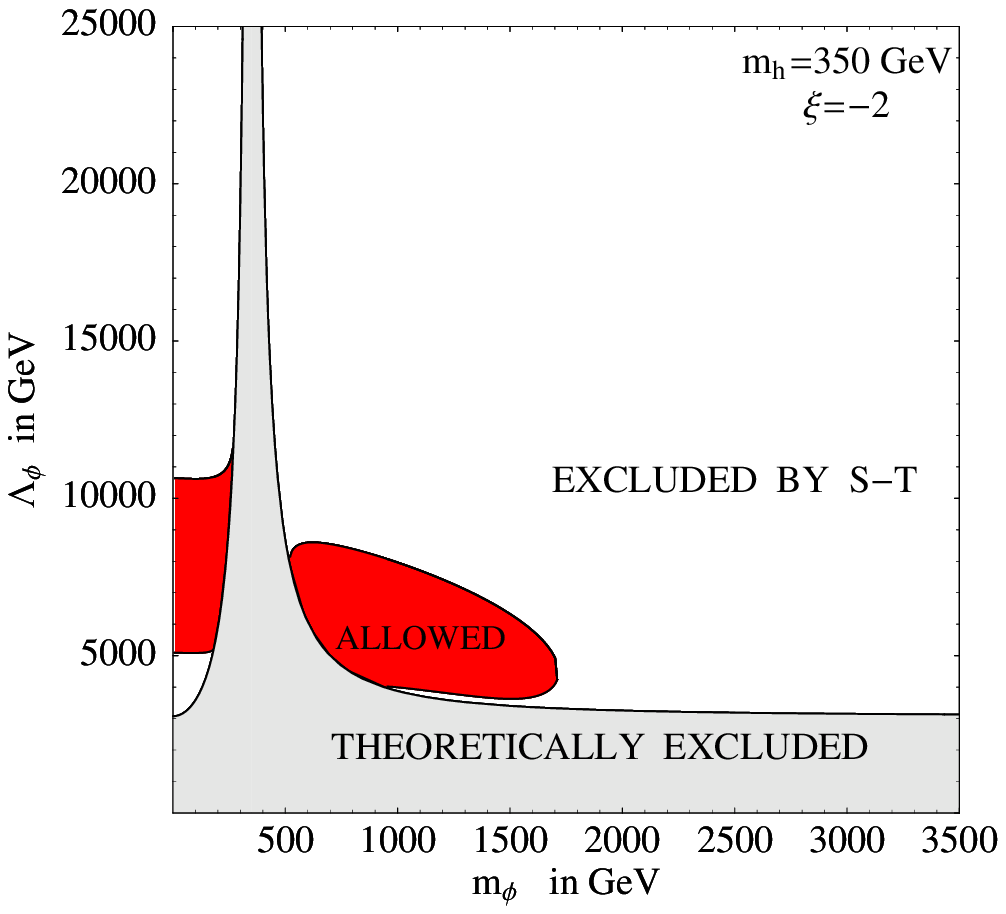}
\includegraphics*[width=6.5cm]{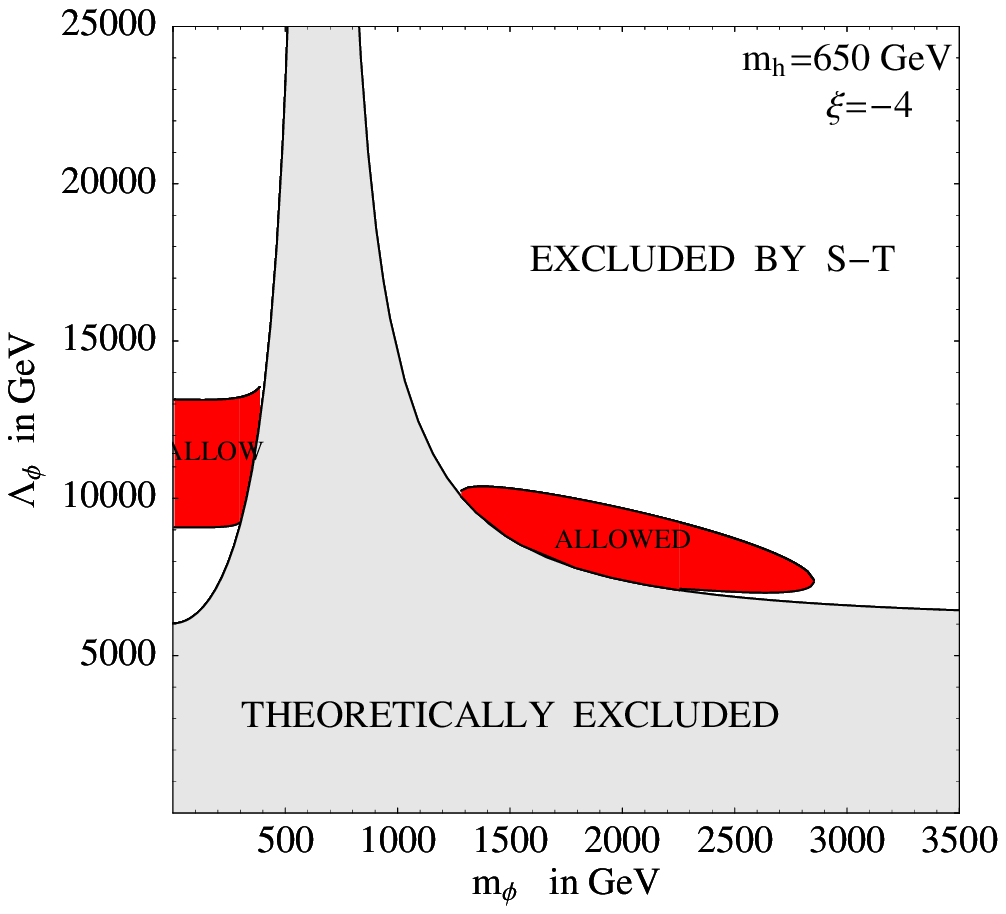}
\caption{Precision electroweak allowed and disallowed regions (at 90\% CL) 
of the radion mass as a function
of $\lphi$, whith fixed values of $\mh$ and $\xi$. In both panels $\eta=1/4$.  
}
\label{mL350650}
\end{figure}

We now consider large scalar masses, and large values of the mixing parameter $\xi$.
In Fig.~\ref{mL350650}, we fix $\mh=350\gev$, $\xi=-2$ (left) and $\mh=650\gev$, $\xi=-4$ (right), 
and show 90\% confidence level contours of allowed regions and disallowed regions in the
$\mphi$-$\lphi$ plane.


The lightly shaded tower in this figure is ``theoretically excluded'' because
it is impossible for both mass eigenstates to exist for the particular
set of input values (one of the fields becomes a ghost).
The darker (red) shaded regions
are allowed by precision electroweak data.
These regions are characterized by $\lphi$ near the theoretically
allowed minimum value and $\mphi$ somewhat near the Higgs mass. We see that
an extraordinarily heavy radion mass compared to the Higgs mass is too
disruptive to the $S$-$T$ fits.  


When increasing the Higgs mass with fixed $\xi$, the large $\mphi$ allowed region tends to disappear. 
But, as shown in the right panel of Fig.~\ref{mL350650} we recover a region compatible with 
precision electroweak data constraints with a larger value of $|\xi|$. Results for $\xi>0$ are similar in nature.

Varying the parameter $\eta$ changes somewhat the form of the allowed regions\footnote{See \cite{pewradion} and references therein for a more complete analysis and collider impact.}, but this does not affect the
general conclusion that a heavy Higgs boson
mass and a heavy radion can both be above the putative Higgs mass
upper limit from precision electroweak data.

\section*{Acknowledgments}
The author would like to thank John Gunion and James Wells for their collaboration on this project.



\begin{thebibliography}{0}



\bibitem{Randall:1999ee}
L.~Randall and R.~Sundrum,
Phys.\ Rev.\ Lett.\  {\bf 83}, 3370 (1999)
[hep-ph/9905221].

\bibitem{Csaki:2000zn}
C.~Csaki, M.~L.~Graesser and G.~D.~Kribs,
Phys.\ Rev.\ D {\bf 63}, 065002 (2001)
[hep-th/0008151].

\bibitem{Kribs:2001ic}
G.~D.~Kribs,
eConf {\bf C010630}, P317 (2001)
[hep-ph/0110242].

\bibitem{Das:2001pn}
P.~Das and U.~Mahanta,
Phys.\ Lett.\ B {\bf 528}, 253 (2002)
[hep-ph/0107162].

\bibitem{Davoudiasl:2000wi}
H.~Davoudiasl, J.~L.~Hewett and T.~G.~Rizzo,
Phys.\ Rev.\ D {\bf 63}, 075004 (2001)
[arXiv:hep-ph/0006041].

\bibitem{pewradion}
J.~F.~Gunion, M.~Toharia and J.~D.~Wells, 
[hep-ph/0311219]. 


\bibitem{Giudice:2000av}
G.~F.~Giudice, R.~Rattazzi and J.~D.~Wells,
Nucl.\ Phys.\ B {\bf 595}, 250 (2001)
[hep-ph/0002178].


\bibitem{Dominici:2002jv}
D.~Dominici, B.~Grzadkowski, J.~F.~Gunion and M.~Toharia,
hep-ph/0206192.








\end{thebibliography}
\end{document}